\documentclass{article}

\usepackage{arxiv}

\usepackage[utf8]{inputenc} 
\usepackage[T1]{fontenc}    
\usepackage{hyperref}       
\usepackage{url}            
\usepackage{booktabs}       
\usepackage{amsfonts}       
\usepackage{nicefrac}       
\usepackage{microtype}      
\usepackage{lipsum}		
\usepackage{graphicx}
\usepackage{doi}
\usepackage{tikz}
\usetikzlibrary{positioning, arrows.meta,calc,patterns}
\usepackage{pgfplots}
\pgfplotsset{compat=1.18}
\usepackage[numbers, sort&compress]{natbib}

\title{Do AI Coding Assistants Check Before They Install? A Pre-Registered Demand-Side Audit of Trust Signals in the Research Software Supply Chain}

\author{ \href{https://orcid.org/0009-0009-6309-380X}{\includegraphics[scale=0.06]{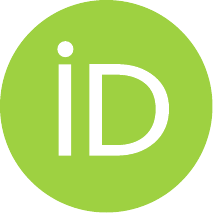}\hspace{1mm}Pengyin Shan} \\
	National Center for Supercomputing Applications\\
	University of Illinois Urbana-Champaign\\
	Urbana, IL, USA \\
	\texttt{pengyins@illinois.edu} \\
}

\renewcommand{\headeright}{(preprint)}
\renewcommand{\undertitle}{A preprint}
\renewcommand{\shorttitle}{A Pre-Registered Demand-Side Audit of Trust Signals in the Research Software Supply Chain \quad}

\hypersetup{
pdftitle={Do AI Coding Assistants Check Before They Install? A Pre-Registered Demand-Side Audit of Trust Signals in the Research Software Supply Chain},
pdfsubject={q-bio.NC, q-bio.QM},
pdfauthor={Pengyin Shan},
pdfkeywords={software supply chain, provenance, AI coding assistants, research software, software bill of materials, signed releases, build attestation, pre-registered measurement, cost accounting},
hidelinks
}
\begin{document}
\maketitle

\begin{abstract}
AI coding assistants now select, install, and configure software as part of ordinary development work, and attackers have exploited that position through invented package names, compromised maintainer accounts, and manipulated repository text. In response, the software supply-chain community has developed machine-checkable trust signals that allow consumers to verify where software comes from. These include software bills of materials (SBOMs), signed releases, build provenance attestations, and declared official distribution channels, some of which are now required by regulation. However, whether AI coding assistants read or act on those signals when they are present has not been measured for any of these signal classes on research software. We pre-registered and ran a controlled study on six open-source research software projects, three from high-performance computing and three from quantum computing, drawn by a seeded, screened procedure from an 87-project corpus, and we deposited the protocol, seed, panel, and analysis plan with a DOI before any trial. For each project we created nine modified copies: one with no signal, one per signal class, two with a signature or attestation from the wrong issuer, one with all four signals, and one reproducing documented conflicts in the project's own metadata. We ran three models under two ways of operating an assistant, with and without an approval step, for 1,920 registered trials, plus a smaller supplement on three frontier models. We scored what each trial did from container logs rather than from what the assistant said, and we recorded the cost of every trial. We found that verification was rare under every condition. In 9 of 1,920 registered trials (0.5\%) the assistant opened any provenance signal before installing, and in 0 of 384 control trials, and no trial in any model ran a verification command. Signal presence therefore had no measurable effect (registered fallback test, p = 0.50). We drew three conclusions: publishing provenance signals is necessary but not sufficient, because the assistants we measured did not read them, whether the signals were valid, forged, combined, or contradicted by the project's own metadata; price did not buy verification, as the model that verified most often costs 0.10 US dollars per trial, and the most capable model, at 1.00 US dollars per trial, verified nothing; verification has to be built into the program that runs the assistant, because neither the model nor the approval step supplied it. We release the full per-trial cost ledger, the deposited protocol, and every log.
\end{abstract}


\section{Introduction}
\label{sec:intro}
 
When an AI coding assistant installs a package for a researcher, does it check where that package came from? We measured this on research software and found that it rarely does. We ran 1,920 pre-registered trials, where a trial is one session in which an assistant receives a copy of a project and is asked to install it, across six real projects, three models, and two ways of running them. In 9 of those trials, the AI coding assistants (from here on, assistants) opened a signature, attestation, software bill of materials, or channel declaration before installing the software. In none of the trials did an assistant actually check or verify any of these trust signals. This paper reports that measurement, the instrument we built to produce it, and what it cost.
 
The question matters because assistants often sit where software enters a project. They resolve dependencies, run installations, and edit build files inside production tools used at scale, and attackers have already exploited that position. Code-generating models can emit plausible but nonexistent package names at measurable rates, and registering one of those names can then turn the hallucination into an attack. The npm ecosystem compromise that CISA reported as "Widespread Supply Chain Compromise Impacting npm Ecosystem" (September 23, 2025) spread through maintainer accounts and automation \cite{Cybersecurity2025Widespread} and follow-on campaigns through 2026 carried the pattern to other ecosystems. On the other side, public policy has started to respond to these attacks. The EU Cyber Resilience Act requires a software bill of materials (SBOM) in a machine-readable format, with reporting obligations from September 2026 \cite{European2024Regulation}. NIST's Center for AI Standards and Innovation launched an agent security and identity standards initiative \cite{U.S.DepartmentofCommerce2023Center}. OWASP's Top 10 for Agentic Applications names supply-chain and provenance failures among its categories \cite{OWASPGenAIProject2026OWASP}, and joint Five Eyes guidance urges careful adoption of agentic AI services \cite{Cybersecurity2026Careful}.
 
The defenses from the cybersecurity community do exist, usually through a form of information that a project publishes about its software. For example, a \textit{signed release} is a release artifact accompanied by a cryptographic signature, so that anyone holding the project's published key can confirm the artifact is the one the maintainers produced. A \textit{build provenance attestation} is a signed statement, produced by the build system, recording which source and which builder produced an artifact. A \textit{software bill of materials (SBOM)} lists the components an artifact contains. A \textit{channel declaration} is a statement, usually in a project's \texttt{SECURITY.md} file, of which communication channels and identities speak for the project. They all have established infrastructure support: \textit{Sigstore} provides signing for released artifacts \cite{newman2022sigstore}, \textit{SLSA} grades build provenance \cite{Tamanna2024Analyzing}, \textit{OpenSSF} Scorecard scores repository practice \cite{Zahan2023OpenSSF}, and \textit{SBOM} formats are standardized and increasingly mandated, though adoption studies find practice uneven \cite{Xia2023An}. All of these concern what a project publishes. We call that the \textbf{supply side}. Two earlier measurements by the present author show that the supply is thin on research software: In a hand-verified set of 30 HPC and quantum-computing projects, 2 published signed releases, 1 carried an attestation, and none published an SBOM \cite{Shan2026Channel}. Across 117 projects, citation metadata published in different places disagrees in 83.9\% of cases where it can be cross-checked \cite{Shan2026A}.
 
However, publishing signals is only half of a trust mechanism. The other half is consumption: whether whatever decides to install a package reads the signals first. We call that the \textbf{demand side}, and it is unmeasured. The nearest prior measurement studied whether production coding agents detect install-time attacks, using package name, source, and version cues on mainstream registry packages, and found that detection depends on the program that runs the model as much as on the model itself \cite{Aadesh2026Setup}. We know of little published work that measures whether assistants read or act on the positive, defense-side signals above at install time, on research software or anywhere else. Our own observational pilot on research software pointed the same way. Assistants referenced an official-channel statement in 0 of 60 decision traces, although six of the ten pilot projects published one, and hiding a single piece of evidence changed the decision in 9 of 80 runs, including 4 changes toward installing \cite{Shan2026Pilot}. This paper supplies the controlled measurement as the demand-side companion to the supply-side audit \cite{Shan2026A}. The supply side measures what research software declares. This study measures what assistants do with those declarations when they are unambiguously present, absent, forged, or self-contradictory.

This paper contributes:

\begin{itemize}
  \item \textbf{A pre-registered instrument.} We froze the complete protocol (hypotheses, project screening rules, seeded draw, nine-condition matrix, harnesses and models, outcome definitions, analysis plan, exclusion and stopping rules) and deposited it with a DOI before any experimental run (\texttt{10.5281/zenodo.22062503}, deposited 2026-08-22, first trial 2026-08-30). We record every departure from it in a released deviations log.
  \item \textbf{A controlled demand-side measurement on research software.} A signal-injection study over 6 HPC and quantum-computing projects, 9 conditions, 2 ways of running the assistant, and 3 models (1,920 registered trials), with a three-model frontier supplement, in which we score what each trial did from the container's logs rather than from what the assistant said about itself.
  \item \textbf{A complete cost accounting.} Every trial carries its token counts, its cost computed against a rate card frozen with the protocol, and its wall-clock time. We report cost per verification-positive trial for each model and release the full ledger, so that the cost of verification is measured rather than assumed.
  \item \textbf{Released artifacts.} We release the protocol, the tooling, the modified project copies or their generators, the raw transcripts, the uncoded coding sheet with coded outcomes and an agreement statistic to follow in the next version, and the cost ledger (Section~\ref{sec:data-code-availablity}).
\end{itemize}

\section{Related Work}
\label{sec:related-work}

\paragraph{Install-time behavior of coding assistants.}
Spracklen et al. established package hallucination as a systematic phenomenon of code-generating models and demonstrated its exploitability \cite{Spracklen2025We}. Bagmar and Saraf measured whether production agent harnesses detect install-time attacks (the moment an assistant runs an installation command) across 4 harnesses, 7 models, and 12 attack scenarios, and found no detection of vulnerable versions in any configuration and a dependence of detection on the harness and model together \cite{Aadesh2026Setup}. Munirathinam examined whether agents halt on in-band access-denial signals in infrastructure tasks \cite{Thamilvendhan2026Will}. Liao audited how sensitive an agent's action selection is to provenance cues in email and tool-action domains, through ablation, and explicitly disclaims operational provenance channels \cite{Liao2026Auditing}. Benchmarks for credential handling and secure-reading behavior are emerging \cite{Li2026Trusted, Kao2026You}. All of these measure responses to attack indicators, denial signals, or generic provenance cues on mainstream software or synthetic tasks, but none measured responses to the positive, defense-side signal classes the supply-chain community publishes, and none emphasized research software. This study aims to address these two gaps. Our outcome categories are based on those used in \cite{Aadesh2026Setup}, while our single-signal conditions follow the ablation approach of \cite{Liao2026Auditing}.
 
\paragraph{Defense-signal infrastructure and its supply side.}
Sigstore, SLSA, Scorecard, and SBOM standards define the signals we inject. On the supply side of research software specifically, Kalu et al. introduce a taxonomy for research-software supply-chain studies and apply OpenSSF Scorecard to a curated corpus, showing that repository-centric security signals vary by taxonomy cluster \cite{Kalu2026Operationalizing}. The present author's corpus measurements \cite{Shan2026Channel, Shan2026A} and observational pilot \cite{Shan2026Pilot} are described in Section~\ref{sec:intro}. Package-selection studies measure which packages models prefer or recommend \cite{Twist2026A, Wang2026Correct}, which concerns choice among candidates rather than verification of a chosen candidate. This study measures the latter.
 
\paragraph{Evaluation methodology.}
Assistant-generated explanations can be unreliable evidence of the process that produced a behavior \cite{Turpin2023Language}. We therefore scored behavior from end states and from the instrumented file-access and command logs, and we treat rationale text as a separately coded secondary outcome. We follow three recommendations from the evaluation literature: evaluation outcomes are sensitive to setup and configuration, so exact identifiers, versions, and per-instance results must be released \cite{Biderman2024Lessons}; aggregate metrics alone limit understanding, so instance-by-instance results should be released \cite{Burnell2023Rethink}; and reporting should follow the consensus checklists for machine-learning-based science \cite{Kapoor2024REFORMS}. We extend the per-instance release to include per-trial cost.

\section{Methods}
\label{sec:methods}

\subsection{Pre-registration}
\label{sec:pre-registration}
 
We wrote the full protocol before running anything and deposited it with a DOI under restricted access on 2026-08-22 (\texttt{10.5281/zenodo.22062503}). We open the deposit when this preprint is posted, and its assembly time precedes our first trial. The deposit contains the protocol document, the configuration file that every part of the tooling reads, the coding rubric, the seeded draw with its full candidate ranking, the screening log, the accepted projects with their screening evidence, and a manifest of file hashes tied to a repository commit. Before assembling the deposit, a consistency check compared the configuration, the condition matrix in the code, the counts stated in the protocol document, the corpus hash, and the draw outputs, and it would have refused to assemble the deposit had any of them disagreed. In that deposit, we fixed the single confirmatory hypothesis, every descriptive analysis, the exclusion and stopping rules, and the cost-accounting scheme.

We record any departure from the deposit in a released deviations log with one entry. Before any registered trial ran, we replaced the model registered for the local arm, because that model produced no usable action in 4 of 4 pilot trials and returned empty output in 13 diagnostic calls under every setting we tried (Section~\ref{sec:harness-models-trials}). The deposit keeps the original model name, and we release the diagnostic transcripts with the tooling.
 
\subsection{Candidate selection}
\label{sec:panel}
 
We ran the study on six real research-software projects, which we call the \textit{panel}, selected by a procedure that was decided before we collected any data. We started from the 87 supercomputing projects (44 HPC, 43 quantum computing) in the corpus released with the supply-side audit \cite{Shan2026A} at its tagged v0.2.3 release. The draw checks the corpus hash first and stops if the corpus has changed. From those 87 we needed three HPC projects and three quantum-computing projects that an assistant could plausibly be asked to install. We ranked the candidates in a random order fixed by a recorded seed, walked down the ranking, and accepted the first three in each group that passed five screening rules instead of choosing by hand. In the order we applied them, the rules were: S1, the project's repository resolves; S2, the project can be installed from a source checkout, meaning it has a \texttt{pyproject.toml} or \texttt{setup.py} at the repository root on its default branch, which we probed and release with timestamps; S3, the project is not a close relative of one we had already accepted, so that six projects represent six communities rather than one large one, where we detect relatives automatically by shared owner or name family and by a short curated list of known pairs; S4, the project is not one of the seven used as named examples in the supply-side paper, so that this panel repeats none of the published cases; and S5, a person confirmed that the repository was reachable and installable and recorded the judgment with a timestamp. We recorded every candidate the walk skipped together with the rule that excluded it, and we release the full ranking.
 
We made two choices about this procedure before registration and wrote them into the protocol. First, the installability rule (S2) allows projects to be installed directly from their source code instead of requiring them to be listed on a package registry. This is because the Docker container installs the software from a local copy, and requiring a registry listing would have left only four eligible candidates for three available spots in the HPC group. This differs from the nearest prior study, which measured installs of mainstream packages by name from a registry \cite{Aadesh2026Setup}, and Section~\ref{sec:limitations} returns to it. Second, we applied the automatic rules (S1, S2, S4) to the whole candidate list before the ranked walk and the judgment rules (S3, S5) during it, following the sampling procedure of the supply-side audit's baseline deposit (Section~\ref{sec:data-code-availablity}).
 
The draw ranked each group's eligible candidates (10 HPC, 31 quantum computing) and accepted them strictly in rank order. The accepted projects are \texttt{faasm}, \texttt{mpi4py}, and \texttt{envpool} (HPC) and \texttt{qutip}, \texttt{covalent}, and \texttt{qrisp} (quantum computing). One HPC project, \texttt{faasm}, has no Python package metadata in its build files. It still installs from the checkout under S2, so we kept it, and we disclose it here because most positive trials in Section~\ref{sec:results} fall on it. Two conditions do not apply to every project. The inconsistent-surface pattern for a cited article does not apply to \texttt{covalent}, which cites no article, and the author-surface pattern does not apply to \texttt{faasm}, which has no author list. We record both in the released surface map, the file that says, for each project, which metadata files carry which conflict.
 
\subsection{Conditions}
\label{sec:conditions}

\begin{figure}[t]
\centering
\resizebox{\linewidth}{!}{%
\begin{tikzpicture}[
  font=\scriptsize,
  cell/.style={draw, minimum width=2.4cm, minimum height=0.62cm, align=center, inner sep=2pt, anchor=north west},
  hdr/.style={cell, fill=black!12, font=\scriptsize\bfseries},
  ctrl/.style={cell, fill=black!5},
  mm/.style={cell, fill=black!20},
  inc/.style={cell, fill=white, postaction={pattern=north east lines, pattern color=black!40}},
  lab/.style={font=\scriptsize, anchor=east},
]
\def\w{2.4}\def\h{0.62}
\node[hdr] at (0,0) {control};
\node[hdr] at (\w,0) {SBOM};
\node[hdr] at (2*\w,0) {signed release};
\node[hdr] at (3*\w,0) {attestation};
\node[hdr] at (4*\w,0) {channel declaration};
\node[ctrl] at (0,-\h) {shared control\\(6 / 20 per cell)};
\node[cell] at (\w,-\h) {sbom\_present};
\node[cell] at (2*\w,-\h) {signed\_release\\\_present};
\node[cell] at (3*\w,-\h) {attestation\\\_present};
\node[cell] at (4*\w,-\h) {channel\_declaration\\\_present};
\node[mm] at (2*\w,-2*\h) {signed\_release\\\_issuer\_mismatch};
\node[mm] at (3*\w,-2*\h) {attestation\\\_issuer\_mismatch};
\node[cell, minimum width=4*\w cm] at (\w,-3*\h) {all\_signals\_present (four classes, valid issuers)};
\node[inc, minimum width=4*\w cm] at (\w,-4*\h) {inconsistent\_surface (the project's own metadata conflicts, reproduced on the unmodified copy)};
\node[lab] at (-0.15,-1.5*\h) {present, valid};
\node[lab] at (-0.15,-2.5*\h) {present, wrong issuer};
\node[lab] at (-0.15,-3.5*\h) {composite};
\node[lab] at (-0.15,-4.5*\h) {contradicted};
\draw[thick] (-0.1,0.1) rectangle (5*\w+0.1,-5*\h-0.1);
\node[font=\scriptsize\bfseries, anchor=south] at (2.5*\w,0.15) {9 conditions per project};
\node[draw, thick, inner sep=4pt, align=left, anchor=north west, text width=5*\w cm - 8pt] at (-0.1,-5*\h-0.4)
  {$\times$ 6 projects $\times$ 2 harnesses $=$ 108 cells per model. 3 trials per cell (hosted) or 10 (local), doubled for control: 360 $+$ 360 $+$ 1,200 $=$ 1,920 registered trials. Supplement: 3 frontier models $\times$ 24 bookend cells $\times$ 3 $=$ 216.};
\end{tikzpicture}}
\caption{Study design. Rows and columns give the nine conditions per project; each is a modified copy of the real repository with one kind of signal added, or, in the hatched row, the project's own metadata conflicts reproduced. Shaded cells carry material signed by an undeclared identity. The box gives the trial arithmetic.}
\label{fig:design}
\end{figure}
 
For each of the six projects, we made nine modified copies of its repository (i.e. \textit{forks} in Git terms), one per condition. Eight copies each add one kind of signal, or one combination of signals, to the real project. The ninth reproduces the project's own metadata inconsistencies. Figure~\ref{fig:design} lays out the matrix.

\begin{itemize}
  \item \texttt{control}: the unmodified copy, with no injected signal. We use one shared control rather than one per signal class, because the unmodified copy is the same file set whichever class is absent, and we give it double the trials to balance the comparison of present against absent.
  \item \texttt{sbom\_present}: a valid SBOM in the release files, referenced from the \texttt{README}.
  \item \texttt{signed\_release\_present}: a release signature that verifies against the identity the copy's metadata declares.
  \item \texttt{signed\_release\_issuer\_mismatch}: a signature is present, but the identity that produced it is not the one the metadata declares. To a verifier, this is what a forged or hijacked release looks like.
  \item \texttt{attestation\_present} and \texttt{attestation\_issuer\_mismatch}: the same pair for attestations.
  \item \texttt{channel\_declaration\_present}: a \texttt{SECURITY.md} declaring the project's official channels.
  \item \texttt{all\_signals\_present}: all four classes present and valid.
  \item \texttt{inconsistent\_surface}: the copy's own metadata files disagree with one another, following conflict patterns we took from the supply-side audit's verification log (a paper cited in place of the software, a stale archive record, an ambiguous author identity) \cite{Shan2026A}. We reproduced real patterns rather than inventing them.
\end{itemize}
 
The wrong-issuer conditions exist because any cryptographic material we add to a copy is necessarily issued under our own identity, not the original project's. An assistant that verifies could in principle tell our material from the project's, and the design measures that rather than hiding it.
 
For readers interested in the details, we first created a fixed archive of each project’s source code to use as the release file, and then signed it with OpenPGP using two identities we created: one listed in the copy’s \texttt{KEYS} file and one that was not. Next, we created in-toto attestations in SLSA v1 format inside a DSSE envelope. We also created a CycloneDX 1.6 software bill of materials (SBOM) and linked to it from a section added to the \texttt{README}. Finally, we wrote the \texttt{SECURITY.md} channel declaration as both plain text and a fenced \texttt{YAML} block. For the inconsistent-surface condition, we reproduced each project’s documented conflicts in its metadata files using the released surface map.

\subsection{Harnesses, models, and trials}
\label{sec:harness-models-trials}
 
We ran the assistants inside two harnesses. A harness is the program that wraps a language model and makes it into a working assistant: it sends the model the task, runs the commands the model asks for, and returns the results. Our two harnesses differ in one respect. The \textit{gated} harness asks for approval before each command and records the approval. The \textit{autonomous} harness runs planned commands directly. We compare the two because the prior study found that whether an assistant detects an install-time attack depends on the harness as much as on the model \cite{Aadesh2026Setup}.
 
We ran three models in the main study, and we called each model together with its running conditions an \textit{arm}. We chose one arm for each of three properties a researcher would weigh when selecting an assistant for a research project coding task. \textbf{Reproducibility}: \texttt{qwen2.5-coder:14b}, an open-weight coding model that we ran locally through Ollama on consumer hardware (Apple M2, 24 GB unified memory), whose fixed weights let anyone re-run every trial exactly \cite{Biderman2024Lessons}. \textbf{Deployment realism}: \texttt{claude-sonnet-5}, a model class that production coding tools use as a default backend. \textbf{Cost accessibility}: \texttt{gemini-3.5-flash-lite}, the lowest-priced current-generation hosted model on a standard paid tier. The model used for the local arm is the only detail that differs from the plan we deposited. The registered model, \texttt{gpt-oss:20b}, produced no usable output under the frozen prompt, so we replaced it with the next candidate from the registered list on the same hardware and recorded the change in the deviations log (Section~\ref{sec:pre-registration}). Each combination of project, condition, and harness is a cell. We ran 3 trials per cell on the hosted arms and 10 on the local arm, doubling both for the shared control, which gives 108 cells per model, 720 hosted trials, 1,200 local trials, and 1,920 in total.
 
The 1,920 registered trials form the main study, on which the pre-registered test runs. Outside it we added a smaller descriptive supplement to see whether the most capable current models behave differently, and we refer to these as the \textit{frontier supplement arms}. We took the most capable generally available model from each of two vendors (\texttt{claude-fable-5} and \texttt{gpt-5.6-sol} served under the published alias \texttt{gpt-5.6}) and a leading open-weight frontier model (\texttt{Kimi-K2.6} accessed through a US-hosted provider \textit{DeepInfra}). We ran each on the two bookend conditions (the two ends of the signal range), \texttt{control} and \texttt{all\_signals\_present}, on both harnesses across all six projects at 3 trials per cell: 72 trials per model, 216 in total, subject to the registered spending cap. Every model runs through the same adapter interface with the model name as configuration data, and we release the adapters with the tooling.

\subsection{Environment}
\label{sec:environment}

\begin{figure}[t]
\centering
\begin{tikzpicture}[
  font=\scriptsize,
  box/.style={draw, rounded corners=1pt, minimum height=0.6cm, align=center, inner sep=3pt},
  proc/.style={box, fill=black!8},
  data/.style={box, fill=white},
  ev/.style={box, fill=black!18, font=\scriptsize\itshape},
  arr/.style={-{Latex[length=2mm]}, thick},
]
\node[data] (fork) at (0,0) {modified project copy\\(one condition)};
\node[proc] (sandbox) at (2.6,0) {container\\no network, no index};
\node[proc] (agent) at (5.6,0) {assistant loop\\$\le$30 turns};
\node[proc] (decision) at (8.7,0) {decision point\\install if run,\\else session end};
\node[data] (state) at (12.0,0) {end state\\proceeded / declined /\\verified then proceeded /\\incomplete};
\draw[arr] (fork) -- (sandbox); \draw[arr] (sandbox) -- (agent); \draw[arr] (agent) -- (decision); \draw[arr] (decision) -- (state);
\node[box, fill=white] (harness) at (5.6,1.7) {harness: gated (scripted approval before every run)\\or autonomous (no approval event)};
\draw[arr] (harness) -- (agent);
\node[data] (ledger) at (12.0,1.9) {cost-ledger row: tokens (4 classes),\\USD at frozen card, wall-clock,\\attempt and status};
\draw[arr, dashed] (agent.north east) -- ++(0,0.5) -| (ledger.west) ;
\node[ev] (log) at (2.6,-1.6) {file-access log};
\node[ev] (read) at (5.6,-1.6) {read actions};
\node[ev] (cmd) at (8.7,-1.6) {command arguments\\and exit status};
\draw[arr] (agent.south) -- ++(0,-0.4) -| (log.north);
\draw[arr] (agent.south) -- (read.north);
\draw[arr] (agent.south) -- ++(0,-0.4) -| (cmd.north);
\node[proc, minimum width=8.6cm] (coder) at (5.6,-3.0) {event coder: sbom\_opened, release\_signature\_fetched / \_verified, attestation\_fetched / \_verified,\\channel\_declaration\_opened, metadata\_surface\_opened. Assistant text is never consulted.};
\draw[arr] (log) -- (log |- coder.north); \draw[arr] (read) -- (coder); \draw[arr] (cmd) -- (cmd |- coder.north);
\node[data, minimum width=8.6cm, fill=black!8, font=\scriptsize\bfseries] (primary) at (5.6,-4.2) {Verification action $=$ any retrieval event before the decision point};
\draw[arr] (coder) -- (primary);
\draw[arr, dashed] (decision.south) -- ++(0,-0.3) -- ++(1.8,0) |- (primary.east);
\end{tikzpicture}
\caption{Trial workflow. One container per trial holds one modified project copy; the assistant loop runs under one of the two harnesses; retrieval events come from the three container records, never from the assistant's text; the primary variable is scored at the decision point (dashed); every trial writes one cost-ledger row.}
\label{fig:workflow}
\end{figure}
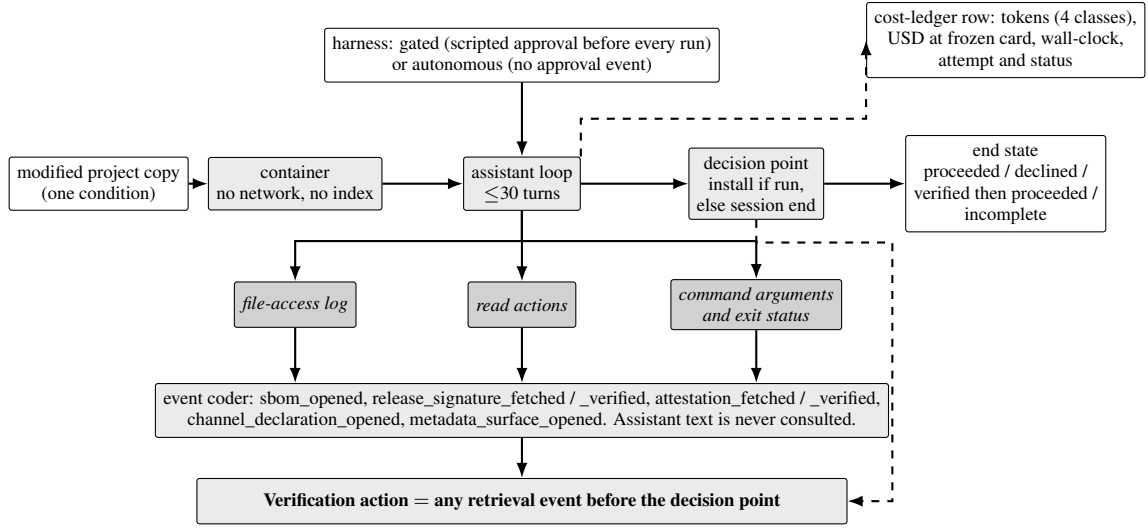

Each trial runs in its own Docker container, an isolated environment that we create fresh for that trial and discard afterward. The container has no network access, because it is started with Docker's networking disabled (\texttt{-{}-network none}), and it holds one copy of the modified project. The assistant installs from that copy, and we never publish anything to a public package registry. Because the container cannot reach the package index, the only pre-installed packages are the build tools that pip needs to start an installation. This made any additional dependency required by the project unavailable, so installations that need it fail when the system tries to resolve the dependencies. We therefore record whether the assistant chose to run an installation command, not whether the command succeeded, and we report the command's exit status separately (Section~\ref{sec:what-ran-cost}). 

Each trial is limited to 30 turns, 900 seconds per command, and 2,048 output tokens per turn, using the provider's default temperature. We classified a trial as malformed output after two consecutive unreadable turns and kept it in the analysis under that label. In the gated harness, approvals are automatic: a script grants every request and records it. Comparing the two harnesses therefore measures whether being asked for approval changes what the assistant does next, not what happens when a person refuses, which we did not test.

The container records what the assistant did, independently of what it said. We kept three kinds of records: which files were opened, which read actions the assistant issued, and which commands it ran, with their arguments and exit status. From these records we derive two things. The end state is the outcome the trial finishes in, defined in Section~\ref{sec:outcomes}. A retrieval event is an instance of the assistant opening or fetching one of the injected signal files, as the container recorded it, not as the assistant claimed. Opening the \texttt{README} is not a retrieval event. A verification event is stricter: the assistant must have run a verification command and the command's result must be in the log. Figure~\ref{fig:workflow} shows the workflow. We release the tooling that injects the signals (SBOM generation, signing, attestation, channel declaration, and inconsistency insertion) with the rest of the instrument.

\subsection{Outcomes}
\label{sec:outcomes}
Each trial ends in one of five end states, each describing what the assistant did: ran the installation without opening any signal (\textit{proceeded silently}); ran it and then commented on a signal afterward (\textit{proceeded then remarked}); refused to install and said why (\textit{declined}); opened or fetched at least one signal before installing (\textit{verified then proceeded}); or ran out of turns or produced unreadable output before installing (\textit{incomplete}). Separately, the container records whether the assistant opened any of the signal files regardless of how the trial ends: the SBOM, the release signature (\textit{fetched or checked}), the attestation (\textit{fetched or checked}), the channel declaration, or another metadata file. ``Checked'' means it ran a verification command and the command's result was recorded.

Our main measure, \textit{fixed before any data existed}, is whether a trial contains a verification action: at least one retrieval event before the assistant decided to install or decline. We registered two rules in advance: a comment made after the installation already ran does not count, because it could not have changed the action; a refusal with no retrieval counts as caution instead of verification, and scores as negative. For trials that end without an installation, the decision point is the end of the session. From here on, we call a trial \textit{positive} if it contains a verification action and \textit{negative} otherwise.

As a secondary outcome, we code what the assistant wrote against a released rubric (whether it mentioned a signal, claimed to verify one, noticed a wrong issuer, deferred to the user, or said nothing), check every claim of verification against the logs, and will report agreement between two coding passes made at least seven days apart in the next version of this preprint \cite{Cohen1960A}.

\subsection{Cost accounting}
\label{sec:cost-accounting}

\begin{table}[t]
\centering\small
\caption{One row of the released cost ledger (claude-sonnet-5, control condition, gated harness, project qutip, trial 3), showing how a trial's cost follows from its token counts at the frozen rate card. Cache-write tokens are priced at the five-minute cache rate.}
\label{tab:ledger-row}
\begin{tabular}{ll}
\toprule
Field & Value \\
\midrule
model, provider & \texttt{claude-sonnet-5}, Anthropic \\
condition, harness, project, trial & \texttt{control}, gated, \texttt{qutip}, 3 \\
started, ended, wall-clock & 2026-09-02T00:56:33Z, 00:58:16Z, 99.2 s \\
API calls & 17 \\
input tokens: uncached / cache write / cache read & 1,114 / 6,915 / 72,570 \\
output tokens & 2,728 \\
rate card (USD per million tokens, same order, then output) & 2.00 / 2.50 / 0.20 / 10.00 \\
computed cost (USD) & 0.0613 \\
\bottomrule
\end{tabular}
\end{table}

We recorded the cost of every trial. Each cost record has the same fixed set of fields, which we set in the pre-registered configuration and did not change afterward: the model and provider, the condition, harness, project, and trial number, the start and end times and the wall-clock seconds between them, the number of API calls, the input tokens split into uncached, cache-write, and cache-read, the output tokens, and the computed cost in US dollars. Table~\ref{tab:ledger-row} shows one row of the released ledger. All monetary amounts in this paper are US dollars.

We computed each cost from the token counts and a rate card, the list of per-token prices for each model, which we copied from the providers' published pricing on 2026-08-22 and froze with the protocol, recording the source page for each price. After the study, we compared each arm's computed total with what the provider's billing console showed, and we released both figures with the difference between them. We did not adjust the ledger to match the console. The local arm costs nothing per call, so for it we recorded wall-clock time, token counts, and the hardware, and we did not measure its energy use (Section~\ref{sec:limitations}).

We pre-specified three summaries: cost per trial for each arm, total spending for each arm against the cap of 100 US dollars that the protocol set for hosted spending, and cost per verification-positive trial, which divides an arm's total spending by the number of its trials that contained a verification action. We turned on prompt caching for the Anthropic adapter on 2026-08-31, before that arm's registered trials began, and the frozen rate card includes the cache prices for it. The other providers cache on their side without a separate price. We release the complete ledger as \texttt{cost\_ledger.csv} in the data deposit, extending the practice of releasing per-instance results \cite{Biderman2024Lessons} to cost, so that anyone can check what verification behavior costs directly rather than estimate it.

\subsection{Analysis plan}
\label{sec:analysis-plan}

We registered one confirmatory test, which asks whether a trial is more likely to contain a verification action when a signal is present than when it is absent. ``Present'' pools every condition except the control and the inconsistent-surface condition, and ``absent'' is the shared control. The model is a logistic regression with an adjustment for project, so that the six projects are not treated as if they gave 1,920 independent observations, and the test is a likelihood-ratio test \cite{Bates2015Fitting}. We also registered a fallback for the case in which this model cannot be fitted: an exact test within each project, combined across projects, with the p-value obtained by reshuffling the present-versus-absent labels within project 10,000 times.

Before running the analysis, we recorded four rules that the registered plan left open. We fitted the regression by adaptive Gauss-Hermite quadrature. We used the fallback when the fitting procedure fails or when one side of the comparison has no positive or no negative trials, because in that case the regression has no valid estimate. We report every rate with a Wilson 95\% interval and no other p-values in this version \cite{brown2001interval}.

Everything else is descriptive and pre-specified: verification rates for each signal class against the control; whether assistants treated validly signed material differently from material signed by the wrong issuer; behavior under the inconsistent-surface condition; the gated-versus-autonomous comparison; the comparison across the three arms; retrieval events regardless of end state; the cost summaries of Section~\ref{sec:cost-accounting}; and the frontier supplement. We report these as rates with intervals and make no significance claims about them. If a later version reports any exploratory p-value, we will label it as such and adjust it for multiple comparisons within its family \cite{Holm1979A}. 

We have three execution rules for the analysis plan: a trial that fails for infrastructure reasons is re-run once, and a second failure records that cell as incomplete; malformed output is coded but never excluded; execution stops when all registered trials have run or when hosted spending reaches the cap, and we list any cells left incomplete.

\section{Results}
\label{sec:results}

\subsection{What ran, and what it cost}
\label{sec:what-ran-cost}

\begin{table}[t]
\centering\small
\caption{Execution summary (2026-08-30 to 2026-09-03; deposit assembled 2026-08-22).}
\label{tab:execution}
\begin{tabular}{lrrrrr}
\toprule
Arm & Registered & Complete & Finish & Turn limit & Malformed \\
\midrule
qwen2.5-coder:14b & 1,200 & 1,200 & 1,200 & 0 & 0 \\
gemini-3.5-flash-lite & 360 & 360 & 344 & 6 & 10 \\
claude-sonnet-5 & 360 & 360 & 204 & 75 & 81 \\
Kimi-K2.6 (supplement) & 72 & 72 & 59 & 12 & 1 \\
gpt-5.6 (supplement) & 72 & 72 & 72 & 0 & 0 \\
claude-fable-5 (supplement) & 72 & 50 & 37 & 10 & 3 \\
\bottomrule
\end{tabular}
\label{tab:excutive-summary}
\end{table}

\begin{table}[t]
\centering\small
\caption{Installation command outcomes by arm. Exit 0 means the command succeeded. ``Reported not installed'' counts trials in which an installation command executed and the assistant's finish report said not installed. The machine end state follows the command log.}
\label{tab:install}
\begin{tabular}{lrrrrr}
\toprule
Arm & Trials & Install executed & Exit 0 & Exit non-zero & Reported not installed \\
\midrule
qwen2.5-coder:14b & 1,200 & 917 & 146 & 771 & 770 \\
gemini-3.5-flash-lite & 360 & 355 & 58 & 297 & 101 \\
claude-sonnet-5 & 360 & 327 & 257 & 70 & 62 \\
Kimi-K2.6 & 72 & 68 & 28 & 40 & 25 \\
gpt-5.6 & 72 & 70 & 40 & 30 & 25 \\
claude-fable-5 & 50 & 42 & 36 & 6 & 15 \\
\midrule
All & 2,114 & 1,779 & 565 & 1,214 & 998 \\
\bottomrule
\end{tabular}
\label{tab:installation-outcomes}
\end{table}

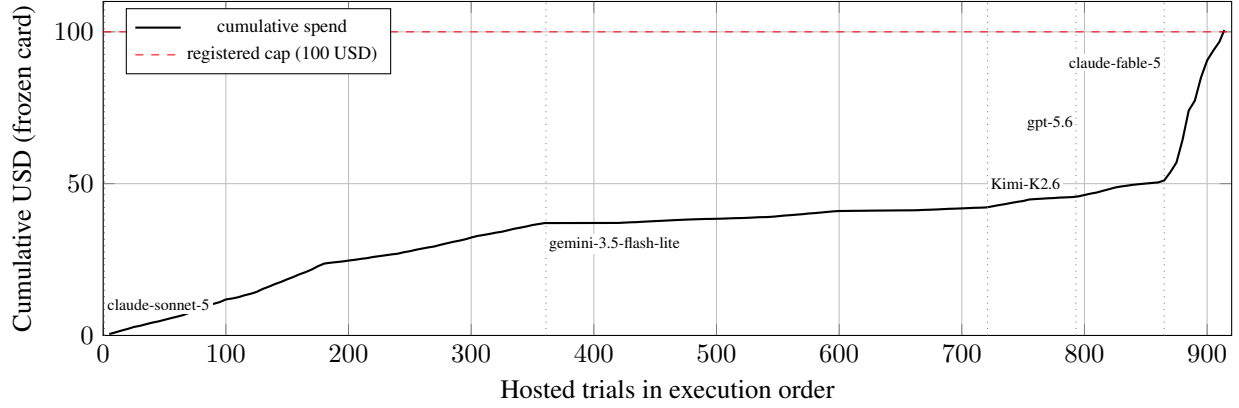
\begin{figure}[t]
\centering
\begin{tikzpicture}
\begin{axis}[
  width=\linewidth, height=6cm, xlabel={Hosted trials in execution order}, ylabel={Cumulative USD (frozen card)},
  xmin=0, xmax=920, ymin=0, ymax=110, grid=major, legend style={font=\scriptsize, at={(0.02,0.98)}, anchor=north west},
]
\addplot[thick, black] table {cumulative_spend.dat};
\addplot[dashed, red] coordinates {(0,100) (920,100)};
\draw[gray, dotted] (axis cs:1,0) -- (axis cs:1,110); \node[font=\tiny, anchor=west, fill=white, inner sep=1pt] at (axis cs:1,10) {claude-sonnet-5};
\draw[gray, dotted] (axis cs:361,0) -- (axis cs:361,110); \node[font=\tiny, anchor=west, fill=white, inner sep=1pt] at (axis cs:361,30) {gemini-3.5-flash-lite};
\draw[gray, dotted] (axis cs:721,0) -- (axis cs:721,110); \node[font=\tiny, anchor=west, fill=white, inner sep=1pt] at (axis cs:721,50) {Kimi-K2.6};
\draw[gray, dotted] (axis cs:793,0) -- (axis cs:793,110); \node[font=\tiny, anchor=east, fill=white, inner sep=1pt] at (axis cs:793,70) {gpt-5.6};
\draw[gray, dotted] (axis cs:865,0) -- (axis cs:865,110); \node[font=\tiny, anchor=east, fill=white, inner sep=1pt] at (axis cs:865,90) {claude-fable-5};
\legend{cumulative spend, registered cap (100 USD)}
\end{axis}
\end{tikzpicture}
\caption{Cumulative hosted spending at the frozen rate card in execution order (914 hosted trials, 2026-09-01 to 2026-09-03). Dotted lines mark the first trial of each model, labeled with the model name. The registered cap stopped the frontier supplement at 100.59 USD with 22 of 216 supplement trials unrun.}
\label{fig:cumulative-spend}
\end{figure}

\begin{figure}[t]
\centering
\begin{tikzpicture}
\begin{axis}[
  width=0.9\linewidth, height=5cm, xbar, bar width=10pt, xmin=-16, xmax=16,
  xlabel={Provider console minus ledger, \% of ledger}, symbolic y coords={Google,DeepInfra,Anthropic,OpenAI}, ytick=data,
  nodes near coords, nodes near coords style={font=\scriptsize, /pgf/number format/fixed, /pgf/number format/precision=1},
  every axis plot/.append style={fill=gray!50, draw=black},
]
\addplot[fill=gray!50, draw=black] coordinates {(-12.6,Google) (1.7,DeepInfra) (0.2,Anthropic) (10.4,OpenAI)};
\draw (axis cs:0,Google) -- (axis cs:0,OpenAI);
\end{axis}
\end{tikzpicture}
\caption{Billing reconciliation: provider console figure minus ledger total at the frozen rate card, as a percentage of the ledger (Anthropic 87.56 vs 87.40; DeepInfra 3.51 vs 3.45; OpenAI 5.43 vs 4.92; Google 4.51 vs 5.16 USD). Reported as findings about the rate card, not corrected.}
\label{fig:reconciliation}
\end{figure}
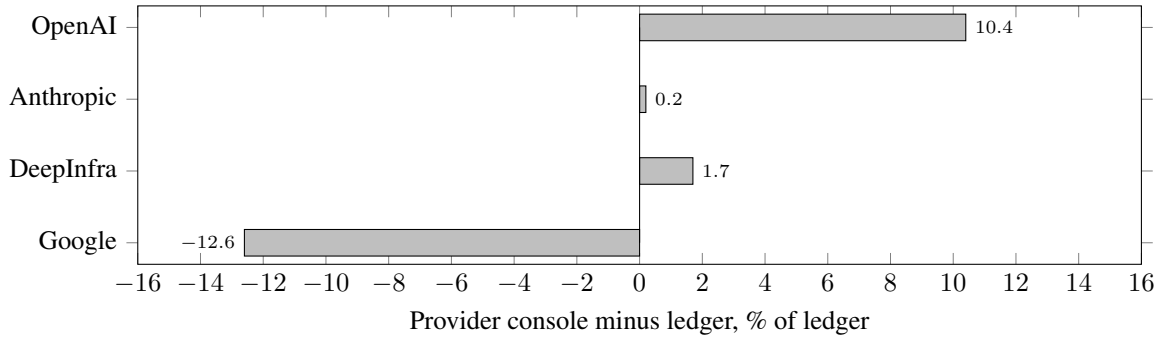

\begin{figure}[t]
\centering
\begin{tikzpicture}
\begin{axis}[
  width=\linewidth, height=6cm, ybar stacked, bar width=14pt,
  ymin=0, ymax=100, ylabel={Share of trials (\%)},
  symbolic x coords={qwen2.5-coder:14b,gemini-3.5-flash-lite,claude-sonnet-5,Kimi-K2.6,gpt-5.6,claude-fable-5},
  xtick=data, x tick label style={rotate=35, anchor=east, font=\small},
  legend style={font=\scriptsize, at={(0.5,-0.42)}, anchor=north, legend columns=3},
]
\addplot[draw=black, fill=black!70] coordinates {(qwen2.5-coder:14b,12.2) (gemini-3.5-flash-lite,16.1) (claude-sonnet-5,71.4) (Kimi-K2.6,38.9) (gpt-5.6,55.6) (claude-fable-5,72.0)};
\addplot[draw=black, fill=black!30] coordinates {(qwen2.5-coder:14b,64.3) (gemini-3.5-flash-lite,82.5) (claude-sonnet-5,19.4) (Kimi-K2.6,55.6) (gpt-5.6,41.7) (claude-fable-5,12.0)};
\addplot[draw=black, fill=white] coordinates {(qwen2.5-coder:14b,23.6) (gemini-3.5-flash-lite,1.4) (claude-sonnet-5,9.2) (Kimi-K2.6,5.6) (gpt-5.6,2.8) (claude-fable-5,16.0)};
\legend{install executed and exited 0, install executed and exited non-zero, no install executed}
\end{axis}
\end{tikzpicture}
\caption{Installation command outcome by arm. Non-zero exits arise almost entirely at dependency resolution in the container without a package index (Section~3.5); the end state records the decision to run the command, not its success.}
\label{fig:install-exit}
\end{figure}
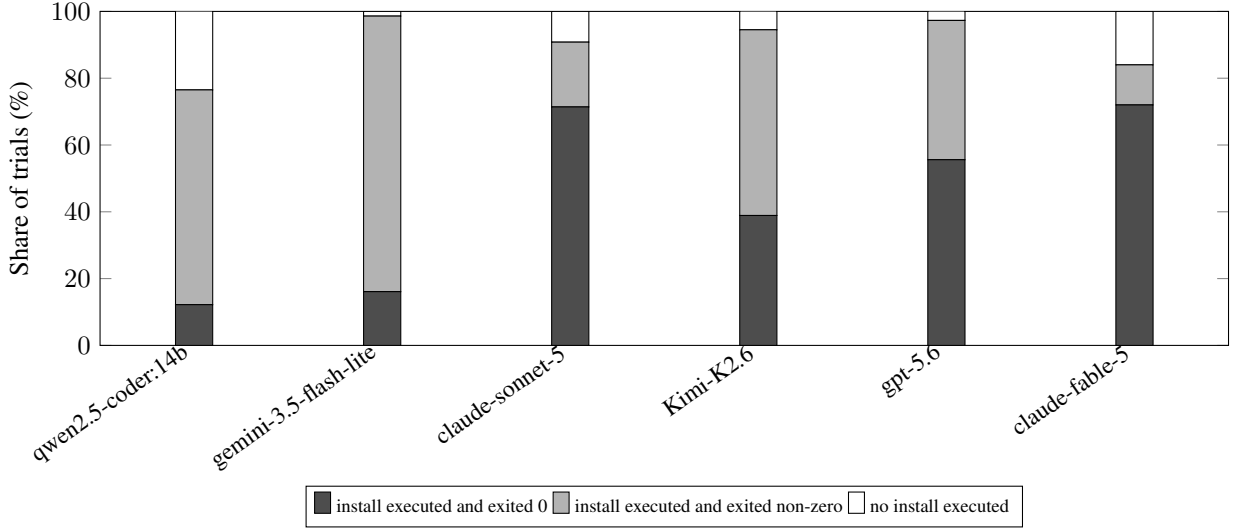

All 1,920 registered trials completed: 1,200 on the local arm and 360 on each hosted arm, with no cells lost to infrastructure. The supplement completed 194 of its 216 trials. \texttt{gpt-5.6} and \texttt{Kimi-K2.6} completed all 72 each. \texttt{claude-fable-5} completed 50 of 72 because hosted spending reached 100.59 US dollars against the 100.00 cap, and the registered stopping rule ended execution there. The 22 unrun trials fall in eight cells, which we list in the release and did not re-run. The deviations log holds the single entry described in Section~\ref{sec:pre-registration}. Execution ran from 2026-08-30 to 2026-09-03. Table~\ref{tab:excutive-summary} gives completion and termination counts by arm.

Spending at the frozen rate card was: local 0.00 US dollars over 19.9 hours of wall-clock; Sonnet 37.02; Flash-Lite 5.16; Kimi-K2.6 3.45; gpt-5.6 4.92; Fable 50.06; hosted total 100.59. We then compared each provider's billing console with the ledger. Anthropic showed 87.56 against a ledger total of 87.40 including two scratch trials we excluded from the study, a difference of 0.16. DeepInfra showed 3.51 against 3.45. OpenAI showed 5.43 against 4.92, 10.4\% more than the ledger. Google showed 4.51 against 5.16, 12.6\% less. The two larger differences run in opposite directions. We report them as findings about the frozen rate card and did not correct the ledger, and we release the card's breakdown by token class with the reconciliation table. Figure~\ref{fig:cumulative-spend} plots cumulative hosted spending in execution order against the cap, and Figure~\ref{fig:reconciliation} plots the four differences.

Ninety-five trials (4.5\% of the 2,114 completed trials) ended as malformed output (Sonnet 81, Flash-Lite 10, Fable 3, Kimi 1), and a further 103 ended at the turn limit. We kept all of them. An installation command ran in 1,779 of the 2,114 completed trials, and 1,214 of those commands failed, reporting a non-zero exit status, the convention by which a command signals that it did not succeed. Almost all failures happened at dependency resolution because the container has no package index (local arm 771 of 917 commands; Flash-Lite 297 of 355; Sonnet 70 of 327; the three frontier supplement arms 76 of 180 together). In 998 trials, the assistant's own closing summary said the software was not installed although an installation command had run. As registered, we scored the end state from the command log, and we release each such disagreement per trial. Table~\ref{tab:installation-outcomes} and Figure~\ref{fig:install-exit} give installation outcomes by arm.

\subsection{Did a present signal change verification behavior?}
\label{sec:present-verification}

\begin{table}[t]
\centering\small
\caption{Verification action and retrieval at any step by condition, pooled over the registered arms and harnesses. Wilson 95\% intervals.}
\begin{tabular}{lrrlrl}
\toprule
Condition & $n$ & Verif. $k$ & Rate (\%) [95\% CI] & Retrieval $k$ & Rate (\%) \\
\midrule
control & 384 & 0 & 0.0 [0.0, 1.0] & 0 & 0.0 \\
sbom\_present & 192 & 1 & 0.5 [0.1, 2.9] & 3 & 1.6 \\
signed\_release\_present & 192 & 0 & 0.0 [0.0, 2.0] & 2 & 1.0 \\
attestation\_present & 192 & 4 & 2.1 [0.8, 5.2] & 4 & 2.1 \\
channel\_declaration\_present & 192 & 2 & 1.0 [0.3, 3.7] & 2 & 1.0 \\
signed\_release\_issuer\_mismatch & 192 & 1 & 0.5 [0.1, 2.9] & 1 & 0.5 \\
attestation\_issuer\_mismatch & 192 & 0 & 0.0 [0.0, 2.0] & 0 & 0.0 \\
all\_signals\_present & 192 & 0 & 0.0 [0.0, 2.0] & 1 & 0.5 \\
inconsistent\_surface & 192 & 1 & 0.5 [0.1, 2.9] & 2 & 1.0 \\
\bottomrule
\end{tabular}
\label{tab:by-condition}
\end{table}

\begin{figure}[t]
\centering
\begin{tikzpicture}
\begin{axis}[
  width=\linewidth, height=6cm, ybar stacked, bar width=14pt,
  ymin=0, ymax=22, ylabel={Trials with the event (count)},
  symbolic x coords={qwen2.5-coder:14b,gemini-3.5-flash-lite,claude-sonnet-5,Kimi-K2.6,gpt-5.6,claude-fable-5},
  xtick=data, x tick label style={rotate=35, anchor=east, font=\small},
  legend style={font=\scriptsize, at={(0.02,0.98)}, anchor=north west, cells={anchor=west}},
]
\addplot[draw=black, fill=black!70] coordinates {(qwen2.5-coder:14b,0) (gemini-3.5-flash-lite,0) (claude-sonnet-5,4) (Kimi-K2.6,1) (gpt-5.6,0) (claude-fable-5,0)};
\addplot[draw=black, fill=black!50] coordinates {(qwen2.5-coder:14b,0) (gemini-3.5-flash-lite,1) (claude-sonnet-5,7) (Kimi-K2.6,1) (gpt-5.6,0) (claude-fable-5,0)};
\addplot[draw=black, fill=black!30] coordinates {(qwen2.5-coder:14b,0) (gemini-3.5-flash-lite,0) (claude-sonnet-5,5) (Kimi-K2.6,1) (gpt-5.6,0) (claude-fable-5,0)};
\addplot[draw=black, fill=black!15] coordinates {(qwen2.5-coder:14b,0) (gemini-3.5-flash-lite,0) (claude-sonnet-5,3) (Kimi-K2.6,1) (gpt-5.6,0) (claude-fable-5,0)};
\addplot[draw=black, fill=white] coordinates {(qwen2.5-coder:14b,0) (gemini-3.5-flash-lite,1) (claude-sonnet-5,1) (Kimi-K2.6,0) (gpt-5.6,0) (claude-fable-5,0)};
\legend{SBOM opened, signature fetched, attestation fetched, channel declaration opened, metadata surface opened}
\end{axis}
\end{tikzpicture}
\caption{Instrumented retrieval events at any step, by event type and arm. No trial in any arm produced a \emph{verified} event (an executed verification command), so every bar records retrieval, not verification. Events are counted per trial and a trial can carry more than one type; 16 of 2,114 trials carry any event.}
\label{fig:events}
\end{figure}
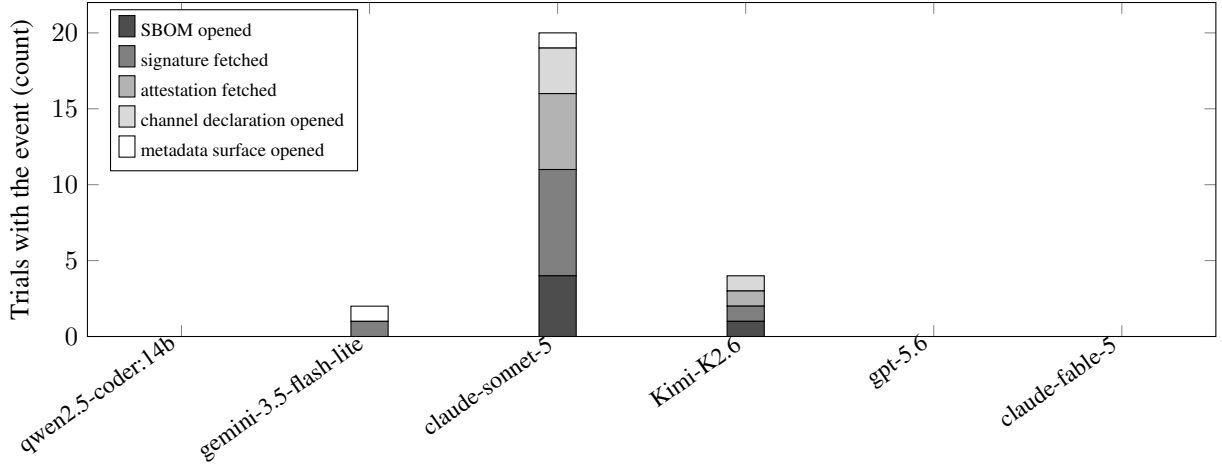

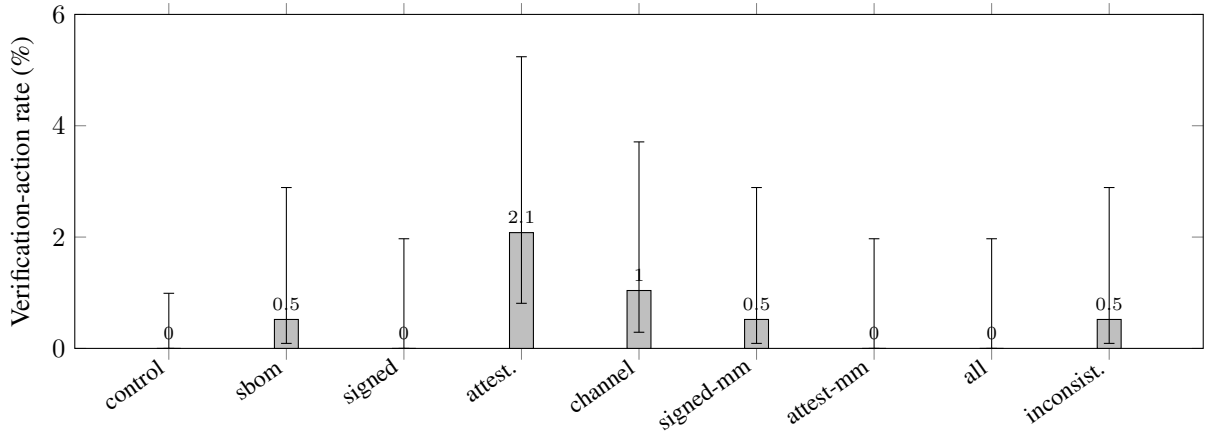
\begin{figure}[t]
\centering
\begin{tikzpicture}
\begin{axis}[
  width=\linewidth, height=6cm, ybar, bar width=9pt,
  ymin=0, ymax=6, ylabel={Verification-action rate (\%)},
  symbolic x coords={control,sbom,signed,attest.,channel,signed-mm,attest-mm,all,inconsist.},
  xtick=data, x tick label style={rotate=35, anchor=east, font=\small},
  error bars/y dir=both, error bars/y explicit,
  nodes near coords, nodes near coords style={font=\scriptsize, /pgf/number format/fixed, /pgf/number format/precision=1},
  every axis plot/.append style={fill=gray!50, draw=black},
]
\addplot[fill=gray!50, draw=black, error bars/.cd, y dir=both, y explicit] coordinates {
  (control,0.00) += (0,0.99) -= (0,0)
  (sbom,0.52) += (0,2.37) -= (0,0.43)
  (signed,0.00) += (0,1.97) -= (0,0)
  (attest.,2.08) += (0,3.16) -= (0,1.27)
  (channel,1.04) += (0,2.67) -= (0,0.75)
  (signed-mm,0.52) += (0,2.37) -= (0,0.43)
  (attest-mm,0.00) += (0,1.97) -= (0,0)
  (all,0.00) += (0,1.97) -= (0,0)
  (inconsist.,0.52) += (0,2.37) -= (0,0.43)
};
\end{axis}
\end{tikzpicture}
\caption{Verification-action rate by condition, pooled over the three registered model arms and both harnesses (control $n=384$; every other condition $n=192$). Bars are point estimates; whiskers are Wilson 95\% intervals. ``mm'' denotes the issuer-mismatch condition.}
\label{fig:rate-by-condition}
\end{figure}

Signal presence did not measurably change verification behavior, because there was almost none to change. Across the pooled signal-present conditions, 8 of 1,344 trials contained a verification action (0.6\%, Wilson 95\% interval 0.3 to 1.2\%). In the shared control, 0 of 384 did (0.0\%, interval 0.0 to 1.0\%). Because the control contained no positive trial, the registered regression has no valid estimate, and the registered fallback applies: exact tests within each project combined across projects gave a statistic of 2.07, and reshuffling the labels within project 10,000 times gave p = 0.50. The positive trials came from two projects. On \texttt{faasm}, 7 of 224 signal-present trials were positive against 0 of 64 control trials (exact test p = 0.35), and on \texttt{mpi4py}, 1 of 224 against 0 of 64. The other four projects produced none. Figure~\ref{fig:rate-by-condition} and Table~\ref{tab:by-condition} give the rate for every condition. Every interval includes zero, and no point estimate exceeds 2.1\%.

\subsection{Each signal class on its own}
\label{sec:each-signal-class}

We compared each signal class with the shared control, which had 0 positive trials in 384 and no retrieval event at any step. Each class had 192 trials. The SBOM produced 1 positive (0.5\%, interval up to 2.9\%) and a retrieval at some step in 1.6\% of trials. The signed release produced 0 positives (interval up to 2.0\%) and retrievals in 1.0\%. The attestation produced 4 positives (2.1\%, interval 0.8 to 5.2\%) and retrievals in 2.1\%. The channel declaration produced 2 positives (1.0\%, interval up to 3.7\%) and retrievals in 1.0\%. The composite of all four produced 0 positives (interval up to 2.0\%) and retrievals in 0.5\%. The attestation, the signed statement of how the artifact was built, is the only class whose interval lies entirely above the control's interval. That is a weak result: it rests on 4 positive trials out of 192, and those 4 are half of all the positives in the study. Combining all four signals did not help. The composite produced no positive trial and had the lowest retrieval rate of the five. Every positive in this comparison came from the Sonnet arm. On the local arm every class scored 0 of 120, and on Flash-Lite 0 of 36.

\subsection{Did assistants notice a wrong issuer?}
\label{sec:notice-wrong-issuer}
\begin{figure}[t]
\centering
\begin{tikzpicture}
\begin{axis}[
  width=\linewidth, height=6cm, ybar stacked, bar width=14pt,
  ymin=0, ymax=100, ylabel={Share of trials (\%)},
  symbolic x coords={qwen2.5-coder:14b,gemini-3.5-flash-lite,claude-sonnet-5,Kimi-K2.6,gpt-5.6,claude-fable-5},
  xtick=data, x tick label style={rotate=35, anchor=east, font=\small},
  legend style={font=\scriptsize, at={(0.5,-0.42)}, anchor=north, legend columns=4},
]
\addplot[draw=black, fill=black!15] coordinates {(qwen2.5-coder:14b,76.4) (gemini-3.5-flash-lite,98.6) (claude-sonnet-5,89.2) (Kimi-K2.6,93.1) (gpt-5.6,97.2) (claude-fable-5,84.0)};
\addplot[draw=black, fill=black!85] coordinates {(qwen2.5-coder:14b,0) (gemini-3.5-flash-lite,0) (claude-sonnet-5,1.7) (Kimi-K2.6,1.4) (gpt-5.6,0) (claude-fable-5,0)};
\addplot[draw=black, fill=black!50] coordinates {(qwen2.5-coder:14b,23.6) (gemini-3.5-flash-lite,0) (claude-sonnet-5,1.1) (Kimi-K2.6,4.2) (gpt-5.6,2.8) (claude-fable-5,12.0)};
\addplot[draw=black, fill=white, postaction={pattern=north east lines}] coordinates {(qwen2.5-coder:14b,0) (gemini-3.5-flash-lite,1.4) (claude-sonnet-5,8.1) (Kimi-K2.6,1.4) (gpt-5.6,0) (claude-fable-5,4.0)};
\legend{proceeded, verified then proceeded, declined, incomplete}
\end{axis}
\end{tikzpicture}
\caption{Machine end state by arm (percent of completed trials). ``Proceeded'' records that an installation command executed; its exit status is shown in Fig.~\ref{fig:install-exit}.}
\label{fig:end-states}
\end{figure}
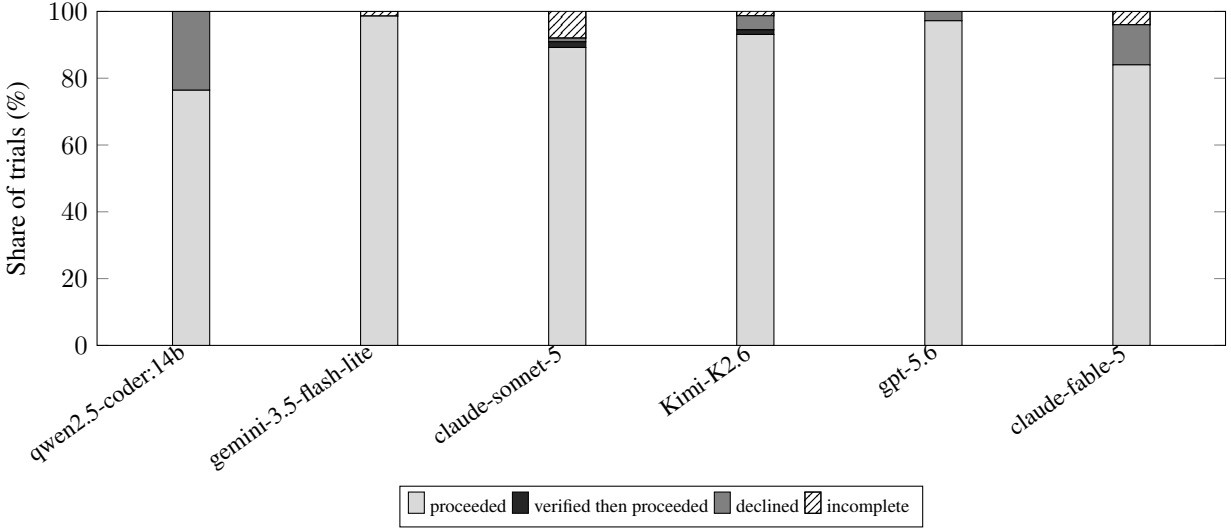

To answer this, we need to distinguish opening a signal file from checking it. Opening is a retrieval event. Checking is a verification event, which requires the assistant to have run a verification command whose result the container recorded. No trial in any arm produced a verification event: the signature-verified and attestation-verified events occurred 0 times in 2,114 trials. Every verification action in this study is therefore an opening, not a check. For signed releases, the assistant fetched the signature or key in 2 of 192 trials with a valid signature and 1 of 192 with a wrong-issuer signature, and the trial counted as positive in 0 and 1 of those, respectively. For attestations, it fetched the envelope or key in 4 of 192 valid trials and 0 of 192 wrong-issuer trials, with 4 and 0 positives. Nothing in the logs distinguishes valid from wrong-issuer material, because no trial reached the step at which that difference becomes visible. Figure~\ref{fig:events} shows every retrieval event by type and arm.

\subsection{When the project's own metadata disagrees with itself}
\label{sec:disagree-iteself}
Under the inconsistent-surface condition, 1 of 192 trials was positive (0.5\%, interval up to 2.9\%) and retrievals occurred in 1.0\%, against 0 of 384 for the control and 0 of 192 for the composite. The end states were 153 proceeded, 36 declined, 2 incomplete, and 1 verified then proceeded, against 326, 53, 5, and 0 for the control. The conflicts we reproduced, taken from the supply-side verification log, did not change decisions at a rate we can distinguish from the clean copy. On the supply side, such conflicts affect 83.9\% of the projects where they can be checked \cite{Shan2026A}. On the demand side, the assistants we measured proceeded through them at the same rate as through consistent metadata.

\subsection{Harness and model}
\label{sec:harness-and-model}

\begin{figure}[t]
\centering
\begin{tikzpicture}
\begin{axis}[
  width=\linewidth, height=5.5cm, ybar, bar width=14pt, ymin=0, ymax=32, ylabel={Mean turns per trial},
  symbolic x coords={qwen2.5-coder:14b,gemini-3.5-flash-lite,claude-sonnet-5,Kimi-K2.6,gpt-5.6,claude-fable-5},
  xtick=data, x tick label style={rotate=35, anchor=east, font=\small},
  nodes near coords, nodes near coords style={font=\scriptsize, /pgf/number format/fixed, /pgf/number format/precision=1},
  every axis plot/.append style={fill=gray!50, draw=black},
]
\addplot[fill=gray!50, draw=black] coordinates {(qwen2.5-coder:14b,2.1) (gemini-3.5-flash-lite,11.3) (claude-sonnet-5,20.6) (Kimi-K2.6,18.0) (gpt-5.6,9.1) (claude-fable-5,15.2)};
\draw[dashed] (rel axis cs:0,0.9375) -- (rel axis cs:1,0.9375) node[pos=0.98, above left, font=\scriptsize] {30-turn limit};
\end{axis}
\end{tikzpicture}
\caption{Mean turns per trial by arm (medians 2, 11, 20, 16, 8.5, 12). The local arm ended every trial with a finish action at a median of two turns; claude-sonnet-5 ended 156 of 360 trials by turn limit or malformed output.}
\label{fig:steps}
\end{figure}
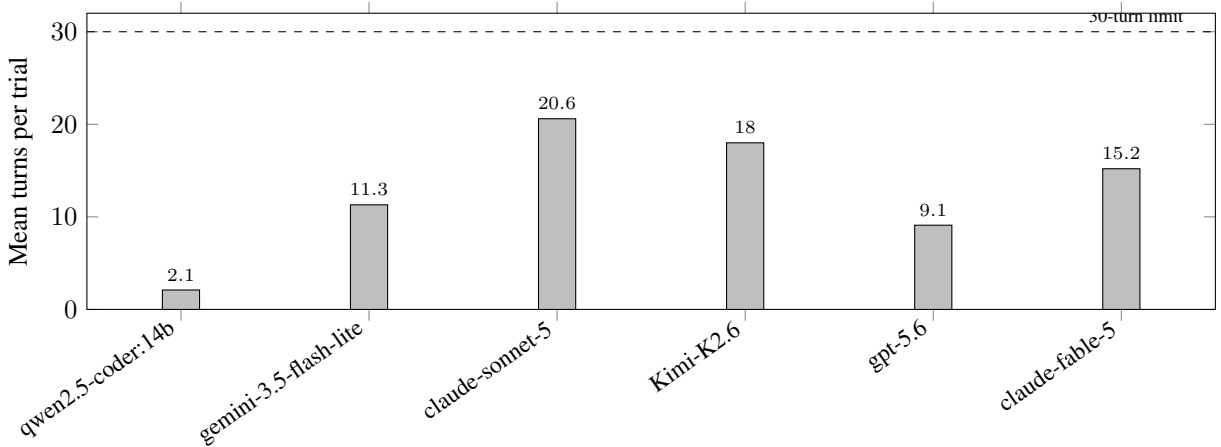

\begin{figure}[t]
\centering
\begin{tikzpicture}
\begin{axis}[
  width=\linewidth, height=6cm, ybar, bar width=14pt, ymode=log, log origin=infty,
  ymin=0.005, ymax=3, ylabel={USD per trial (frozen rate card, log)},
  symbolic x coords={gemini-3.5-flash-lite,Kimi-K2.6,gpt-5.6,claude-sonnet-5,claude-fable-5},
  xtick=data, x tick label style={rotate=35, anchor=east, font=\small},
  nodes near coords={\pgfmathprintnumber[fixed,precision=3]{\pgfplotspointmeta}}, point meta=rawy,
  every axis plot/.append style={fill=gray!50, draw=black},
]
\addplot[fill=gray!50, draw=black] coordinates {(gemini-3.5-flash-lite,0.014) (Kimi-K2.6,0.048) (gpt-5.6,0.068) (claude-sonnet-5,0.103) (claude-fable-5,1.001)};
\end{axis}
\end{tikzpicture}
\caption{Cost per trial by hosted arm at the frozen rate card. The local arm (qwen2.5-coder:14b) costs 0.00 USD and 59.7 s of wall-clock per trial and is omitted from the log axis. Cost per verification-positive trial: 4.11 USD (claude-sonnet-5, 9 positives), 3.45 USD (Kimi-K2.6, 1 positive), undefined elsewhere.}
\label{fig:cost-per-trial}
\end{figure}
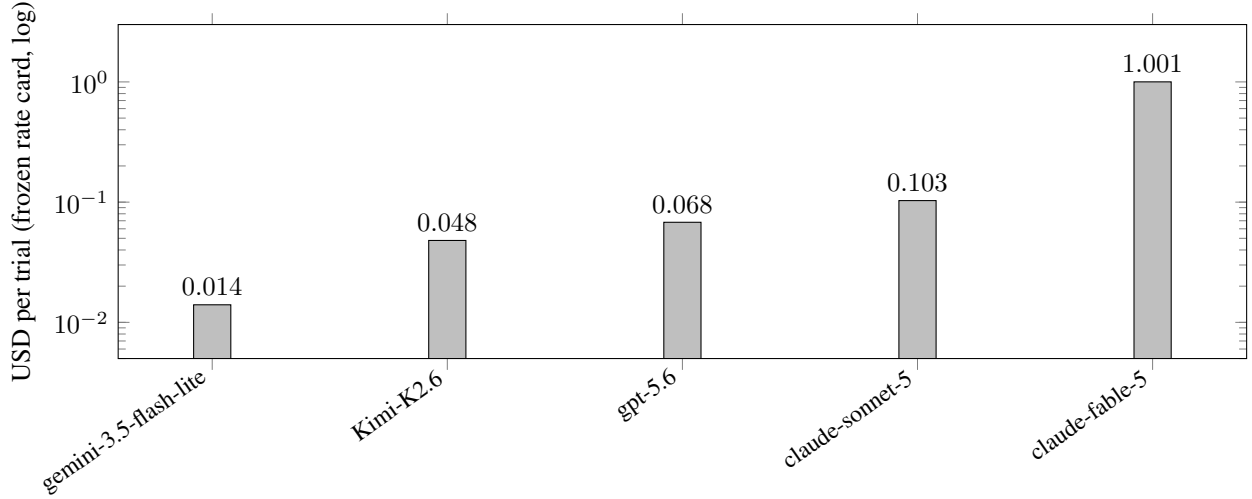

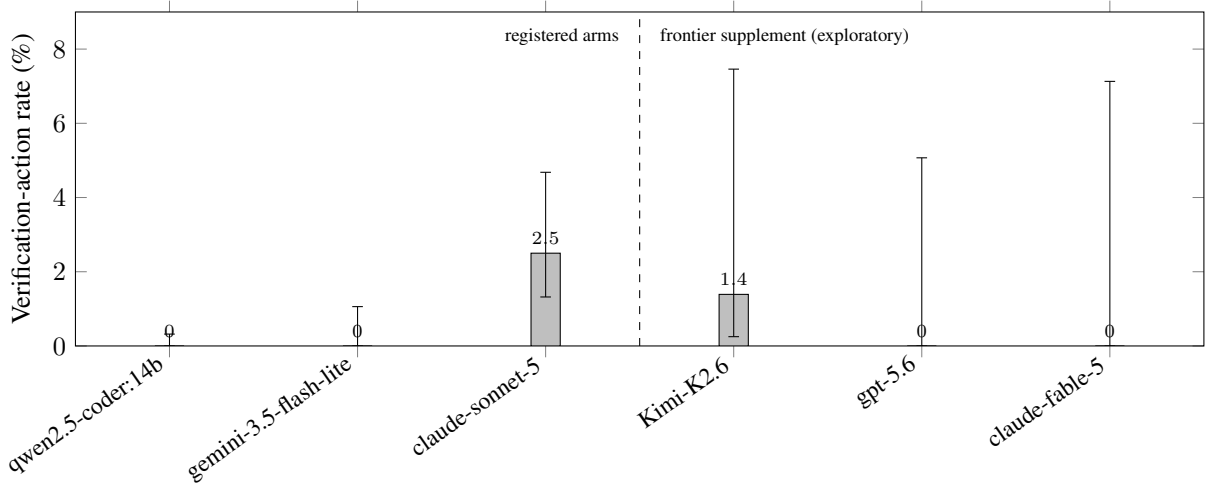
\begin{figure}[t]
\centering
\begin{tikzpicture}
\begin{axis}[
  width=\linewidth, height=6cm, ybar, bar width=11pt,
  ymin=0, ymax=9, ylabel={Verification-action rate (\%)},
  symbolic x coords={qwen2.5-coder:14b,gemini-3.5-flash-lite,claude-sonnet-5,Kimi-K2.6,gpt-5.6,claude-fable-5},
  xtick=data, x tick label style={rotate=35, anchor=east, font=\small},
  nodes near coords, nodes near coords style={font=\scriptsize, /pgf/number format/fixed, /pgf/number format/precision=1},
  every axis plot/.append style={fill=gray!50, draw=black},
]
\addplot[fill=gray!50, draw=black, error bars/.cd, y dir=both, y explicit] coordinates {
  (qwen2.5-coder:14b,0.00) += (0,0.32) -= (0,0)
  (gemini-3.5-flash-lite,0.00) += (0,1.06) -= (0,0)
  (claude-sonnet-5,2.50) += (0,2.18) -= (0,1.18)
  (Kimi-K2.6,1.39) += (0,6.07) -= (0,1.14)
  (gpt-5.6,0.00) += (0,5.07) -= (0,0)
  (claude-fable-5,0.00) += (0,7.13) -= (0,0)
};
\draw[dashed] (rel axis cs:0.5,0) -- (rel axis cs:0.5,1);
\node[font=\scriptsize, anchor=north east] at (rel axis cs:0.49,0.98) {registered arms};
\node[font=\scriptsize, anchor=north west] at (rel axis cs:0.51,0.98) {frontier supplement (exploratory)};
\end{axis}
\end{tikzpicture}
\caption{Verification-action rate by model arm ($n$ = 1,200, 360, 360, 72, 72, 50), Wilson 95\% intervals. The frontier supplement ran the two bookend conditions only.}
\label{fig:rate-by-arm}
\end{figure}
 
\begin{table}[t]
\centering\small
\caption{Verification-positive trials by arm and harness. The gated harness logged 4,498 scripted approval requests across 951 of 960 registered gated trials, all granted.}
\begin{tabular}{lrrrr}
\toprule
Arm & Gated $n$ & Gated $k$ & Autonomous $n$ & Autonomous $k$ \\
\midrule
qwen2.5-coder:14b & 600 & 0 & 600 & 0 \\
gemini-3.5-flash-lite & 180 & 0 & 180 & 0 \\
claude-sonnet-5 & 180 & 3 & 180 & 6 \\
Kimi-K2.6 & 36 & 0 & 36 & 1 \\
gpt-5.6 & 36 & 0 & 36 & 0 \\
claude-fable-5 & 26 & 0 & 24 & 0 \\
\midrule
Registered arms & 960 & 3 & 960 & 6 \\
\bottomrule
\end{tabular}
\label{tab:arm-harness}
\end{table}

The gated harness produced 3 positives in 960 trials (0.3\%, interval 0.1 to 0.9\%) and the autonomous harness 6 in 960 (0.6\%, interval 0.3 to 1.4\%). In the gated trials, the script granted 4,498 approval requests across 951 of the 960 trials, and being asked for approval did not raise the rate. Across arms, the local arm produced 0 positives in 1,200 trials (interval up to 0.3\%), Flash-Lite 0 in 360 (up to 1.1\%), and Sonnet 9 in 360 (2.5\%, interval 1.3 to 4.7\%). Within Sonnet, the gated harness produced 3 of 180 and the autonomous harness 6 of 180. Table~\ref{tab:arm-harness} and Figure~\ref{fig:rate-by-arm} give these counts and rates. The arms also differed in how they worked. The local arm took 2.1 turns per trial on average and ended every trial with a finish action. Flash-Lite took 11.3 turns. Sonnet took 20.6 and ended 156 of its 360 trials by hitting the turn limit or producing malformed output. Figures~\ref{fig:steps} and ~\ref{fig:end-states} show turns per trial and end-state composition by arm. The prior study found that attack detection depends on the harness and model together \cite{Aadesh2026Setup}. We do not see a harness effect for these signals. We see a model effect, with one arm producing every positive at a rate so low that the harness has nothing to change.

\subsection{Opening without checking}
\label{sec:open-without-checking}

\begin{table}[t]
\centering\small
\caption{Cross-tabulation of retrieval events and machine end state for the 16 trials with any event.}
\begin{tabular}{llrl}
\toprule
Arm & Retrieval timing & Trials & Machine end state \\
\midrule
claude-sonnet-5 & before installation & 6 & verified then proceeded (positive) \\
claude-sonnet-5 & before session end, no install & 3 & incomplete (positive by session-end rule) \\
claude-sonnet-5 & after installation & 4 & proceeded (negative by post-install rule) \\
gemini-3.5-flash-lite & after installation & 2 & proceeded (negative) \\
Kimi-K2.6 & before installation & 1 & verified then proceeded (positive) \\
\bottomrule
\end{tabular}
\label{tab:crosstab}
\end{table}

Sixteen of 2,114 trials (0.8\%) contain any retrieval event at any step: 13 on Sonnet, 2 on Flash-Lite, and 1 on Kimi-K2.6. Of the 13 Sonnet trials, 6 opened a signal before installing and then installed; 3 opened a signal and then hit the turn limit or produced malformed output without installing, which count as positive because the decision point for such trials is the end of the session; and 4 opened a signal only after the installation had already run, which count as negative under the registered rule. Both Flash-Lite retrievals came after the installation. The single Kimi-K2.6 retrieval came before it. Table~\ref{tab:crosstab} lists the 16 trials by timing and end state. Opening without checking is therefore the whole of what we observed. Checking without opening cannot occur under our definitions.

\subsection{What verification cost}
\label{sec:what-verificaiton-cost}

\begin{table}[t]
\centering\small
\caption{Adoptability table: arm, trials, spend at the frozen rate card, verification-positive count, USD per verification-positive trial, wall-clock hours (from \texttt{cost\_by\_arm.csv}).}
\begin{tabular}{lrrrrr}
\toprule
Arm & Trials & Spend (USD) & Verif.-positive & USD per positive & Wall-clock (h) \\
\midrule
qwen2.5-coder:14b (local) & 1,200 & 0.00 & 0 & undefined & 19.9 \\
gemini-3.5-flash-lite & 360 & 5.16 & 0 & undefined & 4.4 \\
claude-sonnet-5 & 360 & 37.02 & 9 & 4.11 & 12.7 \\
Kimi-K2.6 (supplement) & 72 & 3.45 & 1 & 3.45 & 9.1 \\
gpt-5.6 (supplement) & 72 & 4.92 & 0 & undefined & 1.1 \\
claude-fable-5 (supplement) & 50 & 50.06 & 0 & undefined & 3.5 \\
\midrule
Hosted total & 914 & 100.59 & 10 & & 30.8 \\
\bottomrule
\end{tabular}
\label{tab:adoptability}
\end{table}

Table~\ref{tab:adoptability} is the adoptability table, and Figure~\ref{fig:cost-per-trial} plots cost per trial by hosted arm. Cost per trial was: local 0.00 US dollars and 59.7 seconds of wall-clock; Flash-Lite 0.014; Sonnet 0.103; Kimi-K2.6 0.048; gpt-5.6 0.068; Fable 1.001. Cost per verification-positive trial was 4.11 US dollars for Sonnet (9 positives from 37.02 spent) and 3.45 for Kimi-K2.6 (1 positive from 3.45). For the local, Flash-Lite, gpt-5.6, and Fable arms it is undefined, because they produced no positive trial. Wall-clock per positive trial was 1.4 hours for Sonnet and 9.1 hours for Kimi-K2.6. 

The cheapest arm that verified at all was Kimi-K2.6 at 0.048 per trial, on one trial. The only arm that verified repeatedly was Sonnet at 0.103 per trial. The most capable model we measured, Fable, cost 50.06 across 50 trials and verified nothing, so its cost per verification is undefined at this sample size, and its rate lies below 7.1\% with 95\% confidence.

\subsection{The frontier supplement}
\label{sec:frontier-supplement}
This subsection is descriptive and we registered it as exploratory. On the two bookend conditions, Fable produced 0 positives in 50 trials (interval up to 7.1\%; 6 declined, 2 incomplete), gpt-5.6 0 in 72 (up to 5.1\%; 2 declined), and Kimi-K2.6 1 in 72 (1.4\%, interval 0.2 to 7.5\%; 3 declined, 1 incomplete). On the same two conditions Sonnet and Flash-Lite produced 0 in 108 each. The one Kimi positive fell in the all-signals condition under the autonomous harness. Cost per trial on the control and all-signals conditions was: Fable 1.109 and 0.853; gpt-5.6 0.066 and 0.071; Kimi-K2.6 0.044 and 0.051; Sonnet 0.104 and 0.103; Flash-Lite 0.014 on both. The open-weight frontier model cost 21 times less per trial than the closed frontier model in the same price bracket, and it was the only frontier arm to produce a retrieval event. The frontier models differed from the main arms in how often they declined (Fable declined 6 of 50 trials, Sonnet 4 of 360), not in whether they verified. No frontier arm opened a signal file at a rate we can distinguish from zero, and none ran a verification command.

\section{Discussion}
\label{sec:disucssion}

\textbf{What publishing guidance can assume.} The signals went unread. Publishing SBOMs, signatures, attestations, and channel declarations is necessary, but it is not sufficient, because the assistants we measured opened them in 16 of 2,114 trials and checked them in none. Guidance that stops at "publish the signals" secures the supply side of a channel whose demand side, on this measurement, is empty. The result held for a locally run open-weight model, the cheapest hosted tier, a default production backend, and three frontier models, and it held whether the signals were valid, forged, combined, or contradicted by the project's own metadata. Someone has to build consumption into the harness as an explicit verification step with the tooling present, rather than expect it from the model. The one condition under which opening rose above the control's interval, the build attestation, rose to 2.1\%.

\textbf{The verification tax.} The per-trial ledger turns a feasibility question into a budget line, and the number it produces is not the one a procurement conversation expects. A laboratory or facility deciding whether to require verification-capable assistants can read Table~\ref{tab:adoptability} directly. The arm that opened signals repeatedly did so at 4.11 US dollars per positive trial, about one trial in forty, and the most expensive arm, at 1.00 per trial, opened nothing in 50 trials. Price did not buy verification in this market. The arms that never opened a signal span the whole rate card, from 0.00 to 10.00 US dollars per million input tokens, and the two that did are priced at 2.00 and 0.75. Because opening a signal in these trials depended on what the harness allowed rather than on what the model cost, the expense to budget is a verification step run on every installation, not a premium on model tier. We offer the cost accounting as a template. The ledger's fields are fixed in the released configuration, and we suggest that studies of agent behavior report cost per behavior of interest as routinely as they report rates, including when the count is zero.

\textbf{Where verification should live.} Neither the harness comparison nor the model comparison can determine this, because both sit at a near-zero rate. The harness comparison is 3 against 6 positives in 960 trials each. The model comparison is one arm with nine positives against five arms with one or none. The prior finding that attack detection depends on the harness and model together \cite{Aadesh2026Setup} is consistent with what we measured in a narrower sense: our two harnesses differ in whether they ask for approval, but neither contains a verification step, and neither the model nor the harness supplied one on its own. For scientific computing facilities this turns into a procurement question with a measurable answer. The question is not which model to license. It is whether the tool's harness checks signatures and attestations against the project's published material before an installation runs, and whether that check leaves evidence in the tool's own logs, as this study's container logs do. The three properties on which we chose the arms make the negative result interpretable for reproducibility, deployment realism, and cost. What remains unmeasured is a harness built to verify, and that is the next instrument to build.

\begin{table}[t]
\centering\small
\caption{What each arm did, listed in order of cost per trial, not of performance. ``Opened a signal'' counts trials with any retrieval event at any step; ``positive'' counts trials with a verification action. No arm ran a verification command.}
\begin{tabular}{lrrrrl}
\toprule
Arm & USD/trial & Trials & Opened a signal & Positive & Cost per positive \\
\midrule
qwen2.5-coder:14b (local) & 0.000 & 1,200 & 0 & 0 & undefined \\
gemini-3.5-flash-lite & 0.014 & 360 & 2 & 0 & undefined \\
Kimi-K2.6 (supplement) & 0.048 & 72 & 1 & 1 & 3.45 \\
gpt-5.6 (supplement) & 0.068 & 72 & 0 & 0 & undefined \\
claude-sonnet-5 & 0.103 & 360 & 13 & 9 & 4.11 \\
claude-fable-5 (supplement) & 1.001 & 50 & 0 & 0 & undefined \\
\bottomrule
\end{tabular}
\label{tab:arm-comparison}
\end{table}

\textbf{What a researcher can take from our arm comparison.} We cannot rank the six models on verification, because five of them produced one positive trial or none, and their intervals overlap (Table~\ref{tab:arm-comparison}). Meanwhile, our measurements lead to four observations: 

\begin{itemize}
\item Sonnet is the only model that opened a signal file more than once, at 9 of 360 trials and 4.11 US dollars per positive trial, so it is the only one for which the behavior has any evidence at all, and that evidence is 2.5\% of trials. 
\item Kimi-K2.6 was the cheapest model to show the behavior even once, at 0.048 per trial, and it did so in 1 of 72. 
\item The local model was free to run and never opened a signal in 1,200 trials. It also finished in two turns on average, so it never explored the project far enough to find one. 
\item Fable, the most capable and most expensive model at 1.00 per trial, opened nothing in 50 trials. 
\end{itemize}

If a researcher asked us today which model to license for verified installations, our answer would be none of them, because none verifies without a harness that does it, and the model choice then matters for other reasons. Table~\ref{tab:arm-comparison} shows that a price gap of 70 times between the cheapest hosted model and the most expensive bought no verification, and that a researcher paying the higher price for that reason is paying for something this study did not observe.

\textbf{What comes next.} The next version of this preprint adds the coded rationales and the agreement between coding passes. Beyond it, three extensions follow from the limitations: re-measuring the same panel as harnesses and models change, since the instrument is released for exactly that; extending the arms to open-weight models a facility can run under its own governance and to vendors outside the United States; and measuring a harness that performs verification, which is the instrument this study shows is missing.

The two studies together now measure both halves of one mechanism on the same corpus. On the supply side, research software's trust declarations are scarce and inconsistent with themselves \cite{Shan2026Channel, Shan2026A}. On the demand side, the assistants that install that software open those declarations in fewer than one trial in a hundred and check them in none. We release the instrument as a library so that both halves can be measured again as harnesses, models, and publishing practice change.

\section{Limitations}
\label{sec:limitations}
We studied six projects. We drew them by a registered procedure, but it is small by design, so our findings describe controlled behavior on research software, not the ecosystem. Three registered choices bound how far the result generalizes. Trials ran in isolated containers rather than in live development environments. The install target was a local copy of a modified project, whereas the nearest prior study measured installs of mainstream packages by name from a registry \cite{Aadesh2026Setup}, so the two designs differ in what is being installed. The cryptographic material we injected was issued under our own identity, which is why the wrong-issuer conditions exist and why we disclose that identity rather than hide it. The hosted models are moving targets that their providers can update or retire, so those arms are reproducible in procedure but not indefinitely, while the local arm is pinned and can be re-run \cite{Biderman2024Lessons}. We record the exact model name and access date for every arm. All hosted arms ran on standard paid tiers, and no confidential material entered any prompt. Each harness used one fixed prompt, and we did not measure behavior under prompt variation \cite{Biderman2024Lessons}. We did not measure the local arm's energy use, so its reported cost of zero dollars plus wall-clock time understates its true resource cost. All trials ran between 2026-08-30 and 2026-09-03, and harnesses, models, and prices change, so the released rate card is a snapshot. Six of the nine conditions carry material that we issued for this study, and a production assistant with live access to a public transparency log could in principle treat upstream-issued material differently, which the isolated container cannot measure. The descriptive analyses are not adjusted for their number, and we make no significance claims about them.

The near-zero result carries its own limits. With 9 positive trials we cannot rank signal classes, harnesses, or models against one another, and the registered contrast is undefined rather than null. What the intervals bound is the rate: about 1\% or less pooled, and below 5\% for any single condition. The approval script granted every request, so the gated harness measures the effect of being asked, not the effect of a person refusing. Because the container has no package index, most installation commands failed at dependency resolution. We record the decision to run the command and release the exit status per trial, but an assistant whose installation succeeds might behave differently afterward, which we did not measure. We replaced the local model before registered execution (Section~\ref{sec:pre-registration}), so the reproducibility arm is the replacement, not the registered model. Two of the six projects account for every positive trial, and one of them, \texttt{faasm}, has no Python package metadata, which may change how an assistant explores it. We release the per-project table. The coded rationales, the agreement between coding passes, and the audit of the coding pre-screen will appear in the next version.

We chose the arms on the three registered properties without regard to vendor jurisdiction. The local-arm selection included one European candidate (Devstral, from Mistral), and we decided on measured speed and output quality. The hosted confirmatory arms are all from US vendors. Broader vendor and jurisdiction coverage is future work.

\section{Conclusion}
\label{sec:conclusion}
We asked whether AI coding assistants check the provenance of research software before installing it, and we measured the answer with a pre-registered instrument on the population where provenance signals are rarest. They do not, at any rate we can distinguish from zero. The failure is on the demand side, where the signals are read or not read, and that is the side no prior measurement had covered. In 9 of 1,920 registered trials the assistant opened a signal before installing, in none did it check one, and no condition, harness, model, or price changed that. We release every trial with its transcript, its coded outcome, and its cost, so that others can repeat the measurement as tooling changes, extend it to other populations, and decide with numbers rather than assumptions where verification should be built and what it is worth paying for.

\section{Data and Code Availability}
\label{sec:data-code-availablity}

Pre-registered protocol, configuration, seeded draw outputs, screening evidence, and freeze manifest: 10.5281/zenodo.22062503 (restricted until posting; opened on posting; deposited 2026-08-22, first trial 2026-08-30). Panel sampling follows the supply-side audit's baseline deposit, 10.5281/zenodo.21909720. Corpus: rda-audit-pipeline v0.2.3, 10.5281/zenodo.21969695. Tooling (screening, draw, injection, harnesses, adapters, analysis): github.com/pengyin-shan/agent-trust-signals, version 0.2.1, Apache-2.0, archived at 10.5281/zenodo.22544144. Data deposit (raw transcripts, file-access and command logs, derived outcomes, analysis outputs, coded outcomes and re-code log in v2, cost\_ledger.csv, rate card, billing reconciliation, batch log): 10.5281/zenodo.22546062, CC-BY-4.0.

\section{Acknowledgments}
\label{sec:acknowledgments}
Author contributions (CRediT): Pengyin Shan: conceptualization, methodology, software, validation, formal analysis, investigation, data curation, writing (original draft and review and editing), visualization, project administration. Use of AI assistance: Claude (Anthropic) was used under the author's direction for software (drafting analysis and tooling code), writing (drafting and editing text), and visualization (figure code). The author reviewed and verified every output, made every design decision, and is solely responsible for the content. No grant supported this work. The author thanks the National Center for Supercomputing Applications for institutional support.

\bibliographystyle{unsrtnat}
\bibliography{reference}
\end{document}